\documentclass[aps,prx,preprint,showpacs]{revtex4-2}
\usepackage{graphicx}
\usepackage{helvet}
\usepackage{amsmath,amssymb}
\usepackage{bm}
\usepackage{color}

\begin{document}

\title{Mott-driven BEC-BCS crossover in a doped spin liquid candidate, $\kappa$-(BEDT-TTF)$_{4}$Hg$_{2.89}$Br$_{8}$}

\author{Y. Suzuki$^{1,\dagger}$}
\author{K. Wakamatsu$^{1,\dagger}$}
\author{J. Ibuka$^{1}$ }
\author{H. Oike$^{1}$ }
\author{T. Fujii$^{2}$ }
\author{K. Miyagawa$^{1}$ }
\author{H. Taniguchi$^{3}$ }
\author{K. Kanoda$^{1,\ast}$ }

\affiliation{
$^{1}$Department of Applied Physics, University of Tokyo, Bunkyo-ku, Tokyo, 113-8656, Japan\\}
\affiliation{
$^{2}$Cryogenic Research Center, University of Tokyo, Bunkyo-ku, Tokyo, 113-0032, Japan\\}
\affiliation{
$^{3}$Graduate School of Science and Engineering, Saitama University, Saitama 338-8570, Japan\\}

\date{\today}

\begin{abstract}
	The pairing of interacting fermions leading to superﬂuidity has two limiting regimes: the Bardeen-Cooper-Schrieﬀer (BCS) scheme for weakly interacting degenerate fermions and the Bose-Einstein condensation (BEC) of bosonic pairs of strongly interacting fermions. 
While the superconductivity that emerges in most metallic systems is the BCS-like electron pairing, strongly correlated electrons with poor Fermi liquidity can condense into the unconventional BEC-like pairs. 
Quantum spin liquids harbor extraordinary spin correlation free from order and the superconductivity that possibly emerges by carrier doping of the spin liquids is expected to have a peculiar pairing nature. 
The present study experimentally explores the nature of the pairing condensate in a doped spin-liquid candidate material and under varying pressure, which changes the electron-electron Coulombic interactions across the Mott critical value in the system. 
The transport measurements reveal that the superconductivity at low pressures is a BEC-like condensate from a non-Fermi liquid and crosses over to a BCS-like condensate from a Fermi liquid at high pressures. 
The Nernst-eﬀect measurements distinctively illustrate the two regimes of the pairing in terms of its robustness to the magnetic ﬁeld. 
The present Mott tuning of the BEC-BCS crossover can be compared to the Feshbach tuning of the BEC-BCS crossover of fermionic cold atoms.

\end{abstract}


\maketitle

\section{Introduction}

Strong interactions between itinerant electrons in solids cause peculiar correlations or organizations among the electrons beyond the Fermi-liquid framework \cite{Imada_1998}. 
Superconductivity emerging in these circumstances can be outside the celebrated Bardeen-Cooper-Schrieffer (BCS) framework, which assumes an instability of Fermi surfaces formed by degenerate fermions with well-deﬁned momenta \cite{Bardeen_1957}, in that the strong correlation of electrons makes their momenta ill-deﬁned and the electron pairing in momentum space, the central idea of BCS theory, is less pertinent. 
In this case, superconductivity is of the Bose-Einstein condensation (BEC) type, as the size of Cooper pairs is as small as their mean distance, or equivalently, the interaction energy is comparable to the Fermi energy. 
Several unconventional superconductors have been discussed in the framework of a BEC scenario \cite{Emery_1995, Uemura_1991, Kasahara_2014, Hashimoto_2020, Nakagawa_2021}. 
More broadly, the superfluidity of cold fermionic atoms is known to be controllable between the BEC and BCS regimes by the magnetic-field tuning of the interatomic interaction through the Feshbach resonance \cite{Regal_2004, Zwierlein_2004}. 
The relation between the pairing nature and the interparticle correlation is a seminal issue in the instability of interacting fermions at large \cite{Chen_2005, Randeria_2014}. 
Thus, it is expected to identify interacting systems having novel types of correlation and explore the nature of the pairing condensate therein under variation in the interaction in a controllable manner.

Quantum spin liquids emerging on geometrically frustrated lattices are argued to possess peculiar spin correlations possibly linked to fractionalization, entanglement and/or chirality in magnetic excitations \cite{Zhou_2017, Savary_2017, Broholm_2020}. 
The superconductivity possibly brought about by doping against such anomalous magnetic state \cite{Lee_2008, Jiang_arxiv, Ogata_2003, Anderson_1987, Zaanen_2011, Powell_2011, Yokoyama_2013, Gull_2013, Lee_2006} is expected to have unconventional pairing nature out of the BCS framework; however, it has yet to be explored experimentally. 
In the present study, we address this issue by working with a doped spin liquid candidate under pressure variation, and reveal that superconductivity in this system exhibits a novel type Mottness-controlled BEC-BCS crossover.

\begin{figure}
	\includegraphics[width=16cm]{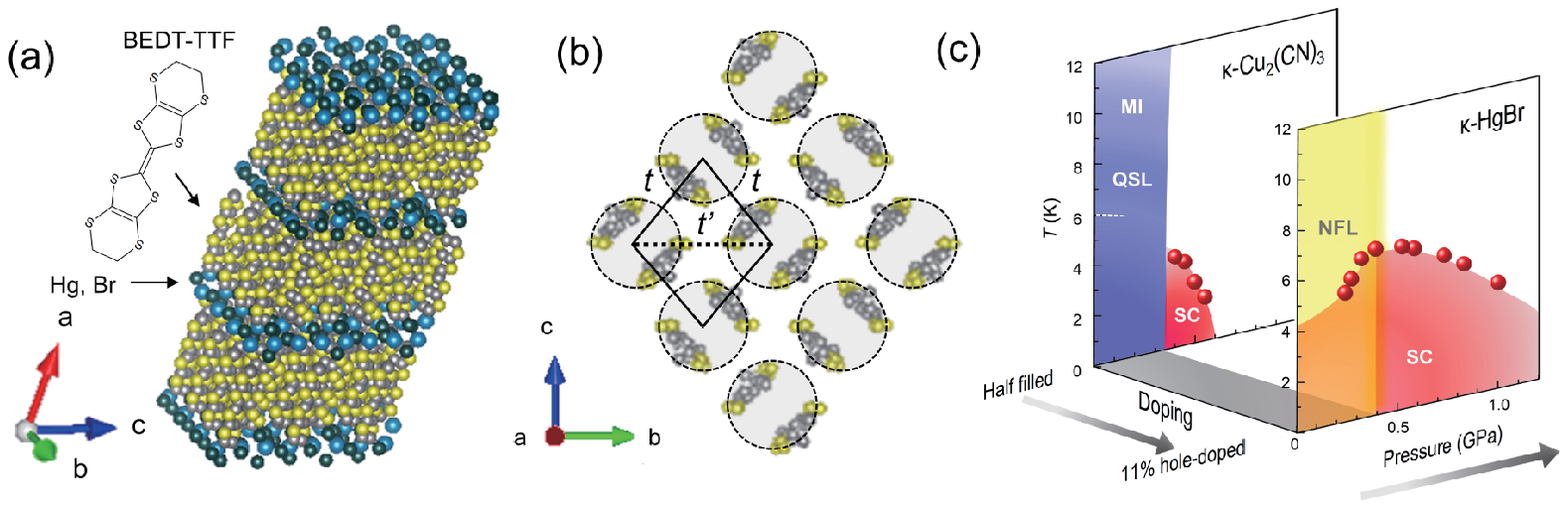}
	\caption{\label{Fig1} (Color online) Structure and phase diagram of $\kappa$-(BEDT-TTF)$_{4}$Hg$_{2.89}$Br$_{8}$. 
(a) Layered structure consisting of BEDT-TTF conducting layers and Hg$_{2.89}$Br$_{8}$ insulating layers. BEDT-TTF denotes bis(ethylenedithio)tetrathiafulvalene. 
The nonstoichiometry in the Hg composition comes from the fixed incommensurability between the Hg and BEDT-TTF lattices originating from material chemistry, not from extrinsic defects. 
(b) In-plane structure of the conducting layers. BEDT-TTF molecules form dimers, which constitute a nearly isotropic isosceles triangular lattice with the anisotropy of transfer integrals, $t^{\prime}$/$t$, close to unity. (c) Pressure-temperature phase diagrams of $\kappa$-(BEDT-TTF)$_{4}$Hg$_{2.89}$Br$_{8}$ and a quantum spin liquid candidate, $\kappa$-(BEDT-TTF)$_{2}$Cu$_{2}$(CN)$_{3}$. 
The red area indicates the superconducting phase. 
The blue area in $\kappa$-(BEDT-TTF)$_{4}$Hg$_{2.89}$Br$_{8}$ depicts the non-Fermi liquid region (corresponding to the yellow region in Fig. \ref{Fig2}(c)). 
The purple area in $\kappa$-(BEDT-TTF)$_{2}$Cu$_{2}$(CN)$_{3}$ depicts the Mott insulating spin liquid phase \cite{Kurosaki_2005, Furukawa_2018}. 
$T_{\textrm{c}}$ in $\kappa$-(BEDT-TTF)$_{2}$Cu$_{2}$(CN)$_{3}$ is taken from Ref. \cite{Kurosaki_2005}, and $T_{\textrm{c}}$ in $\kappa$-(BEDT-TTF)$_{4}$Hg$_{2.89}$Br$_{8}$ corresponds to the midpoint of the resistive transition observed in this study. MI, QSL, NFL and SC represent the Mott insulator, quantum spin liquid, non-Fermi liquid, and superconductivity, respectively. 
The white broken line indicates the location of the so-called 6 K-anomaly observed in $\kappa$-(BEDT-TTF)$_{2}$Cu$_{2}$(CN)$_{3}$ \cite{Zhou_2017, Miksch_2021, Kurosaki_2005}, a possible indication of an instability of QSL.
}
\end{figure}

The material studied is a layered organic superconductor, $\kappa$-(BEDT-TTF)$_{4}$Hg$_{2.89}$Br$_{8}$ ($\kappa$-HgBr hereafter), in which dimers of BEDT-TTF molecules constitute a nearly perfect isotropic triangular lattice, separated by insulating Hg$_{2.89}$Br$_{8}$ layers (Figs. \ref{Fig1}(a) and \ref{Fig1}(b)) \cite{Lyubovskaya_1987, SM}. 
The nonstoichiometry of Hg arises from its sublattice incommensurate with the host lattice formed by BEDT-TTF and Br, and the composition of Hg is reproducibly given by the mismatch between the two lattices precisely determined by x-ray diffraction \cite{Lyubovskaya_1987, SM}.
The triangular lattice structure is almost identical to that of the Mott-insulating spin liquid candidate $\kappa$-(BEDT-TTF)$_{2}$Cu$_{2}$(CN)$_{3}$ ($\kappa$-Cu$_{2}$(CN)$_{3}$ hereafter), in which the spin-1/2 localized on each BEDT-TTF dimer interacts with its neighbors through an antiferromagnetic exchange energy $J$ of 250 K but does not order down to 30 mK \cite{Shimizu_2003, Pratt_2011, Yamashita_2008}.
A distinction of $\kappa$-HgBr from $\kappa$-Cu$_{2}$(CN)$_{3}$ is that the nonstoichiometric Hg ions give 11$\%$ hole doping to the Mott insulator, resulting in a metallic state \cite{SM, Oike_2015, Hebert_2015}. 
Surprisingly, the $J$-scaled spin susceptibility, namely, $J\chi_{\textrm{spin}}$, versus $T$/$J\chi_{\textrm{spin}}$, ($J$ = 140 and 250 K for $\kappa$-HgBr and $\kappa$-Cu$_{2}$(CN)$_{3}$, respectively) nearly coincide between $\kappa$-HgBr and $\kappa$-Cu$_{2}$(CN)$_{3}$, and are well reproduced by the triangular-lattice antiferromagnetic Heisenberg model, suggesting that $\kappa$-HgBr hosts a QSL in the spin sector while remaining metallic in the charge sector \cite{Oike_2017}. 
Thus, $\kappa$-HgBr is a possible realization of the long sought-after doped QSL \cite{Broholm_2020}. 
According to the transport study \cite{Oike_2015, Taniguchi_2007}, the metallic state at low pressures is a non-Fermi liquid (NFL), and the doped holes, not particles of the entire band, carry mobile charges. 
Superconductivity in such a doped QSL candidate \cite{Lyubovskaya_1987, Oike_2015}, possibly appearing in place of the low-temperature instability of the QSL \cite{Miksch_2021}, is of profound interest \cite{Lee_2008, Jiang_arxiv, Ogata_2003, Anderson_1987, Zaanen_2011, Powell_2011, Yokoyama_2013, Gull_2013, Lee_2006}.

Pressure is an effective control parameter of electron-electron interactions in molecular materials, which are highly compressible in general \cite{Kagawa_2005}. 
The transport measurements indicate that when $\kappa$-HgBr is pressurized in excess of 5 kbar, all band particles become mobile \cite{Oike_2015}, and concomitantly, the NFL at low pressures changes to a Fermi liquid (FL). 
Comparative phase diagrams (Fig. \ref{Fig1}(c)) illustrate that the Mott-insulating QSL phase in $\kappa$-Cu$_{2}$(CN)$_{3}$ is replaced by the NFL in $\kappa$-HgBr \cite{Oike_2015, Kurosaki_2005, Furukawa_2018}. 
Remarkably, $T_{\textrm{c}}$ forms a dome around the NFL-FL crossover pressure \cite{Oike_2015, Taniguchi_2007}. In the present work, we investigate the pressure evolution of the nature of electron pairing that occurs in a doped QSL candidate $\kappa$-HgBr by electron transport and Nernst effect measurements. 
The results demonstrate the pressure-tuned BEC-BCS crossover and clarify its relation to the Fermi liquidity above $T_{\textrm{c}}$.

\section{RESULTS and DISCUSSION}
\subsection{Crossover from non-Fermi liquid to Fermi liquid}
\begin{figure}
	\includegraphics[width=15cm]{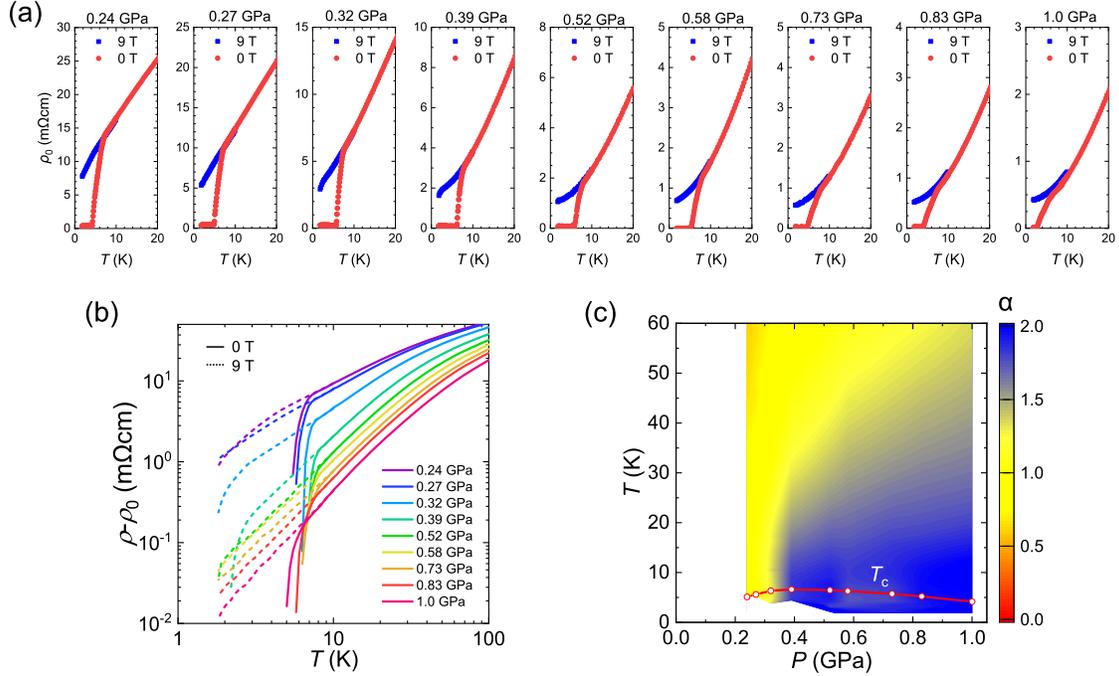}
	\caption{\label{Fig2} (Color online) Temperature dependence of resistivity under pressures. 
(a) Temperature variation of the in-plane resistivity $\rho$($T$) at 0 T and a perpendicular field of 9 T. 
(b) Logarithmic plot of the temperature-dependent part of resistivity, $\rho$($T$) - $\rho_{0}$, with $\rho_{0}$ being the residual resistivity (see main text for its determination). 
The solid and dashed lines indicate the resistivity at $H$ = 0 and 9 T, respectively. 
(c) Contour plot of the exponent $\alpha$ in fits of the form $AT^{\alpha}$ to the temperature-dependent part of the experimental in-plane resistivity $\rho$($T$) - $\rho_{0}$; $\alpha$ is determined by $\alpha$($T$) $\equiv$ $d \log$($\rho$($T$) - $\rho_{0}$)/$d \log$$T$. The yellow region ($\alpha$ $\approx$ 1) corresponds to a non-Fermi-liquid (NFL) phase, and the blue region ($\alpha$ $\approx$ 2) corresponds to the Fermi-liquid (FL) phase.
}
\end{figure}

To precisely associate the nature of electron pairing with that of the normal-state fluidity above $T_{\textrm{c}}$, we performed resistivity measurements under finely incremented pressures \cite{SM}. 
To avoid resistivity jumps due to microcracking of the crystal encountered in the ambient-pressure measurements, we had to apply finite pressures to obtain jump-free resistivity data. 
As seen in Fig. \ref{Fig2}(a), the temperature dependence of the in-plane resistivity $\rho$($T$) varies systematically with pressure; the linear temperature dependence changes into superlinear dependences at elevated pressures. 
Fitting the form of $\rho_{\textrm{n}}$($T$)=$\rho_{0}$+$AT^{\alpha}$ to the normal-state resistivity$\rho$($T$) below 10 K measured under 9 T yields the value of the residual resistivity $\rho_{0}$ for each pressure. The logarithmic plot of ($\rho$($T$) - $\rho_{0}$) versus $T$ (Fig. \ref{Fig2}(b)) highlights the pressure evolution of the temperature-dependent part of resistivity. The slopes of these curves, given by $\alpha$($T$) $\equiv$ $d\log$($\rho$($T$) - $\rho_{0}$)/$d\log$$T$, are defined as the exponent at temperature $T$. 
The contour plot of the deduced $\alpha$($T$) values on the pressure-temperature plane (Fig. \ref{Fig2}(c)) shows that at low temperatures, $\alpha$($T$)  is approximately unity before the pressure reaches 0.4 GPa and sharply switches to the value of 2 at higher pressures, demonstrating that an NFL crosses over to an FL at approximately 0.35-0.40 GPa. 
This finding reproduces the previous results \cite{Oike_2015} and additionally reveals that the crossover is so sharp, as theoretically suggested \cite{Tocchio_2013}, that it is reminiscent of a quantum phase transition \cite{Sondhi_1997, Sachdev_1999, Gegenwart_2008, Furukawa_2015}. 
This sharp crossover is regarded as a correspondence to the Mott transition in $\kappa$-Cu$_{2}$(CN)$_{3}$, as implied by Fig. \ref{Fig1}(c); at the crossover, the interaction strength crosses a Mott critical value at which the double occupancies of holes on a site, namely, Mottness, drastically changes \cite{Oike_2015}, as theoretically suggested \cite{Hebert_2015}. 
Furthermore, the NFL state does not only occur at the quantum critical point but extends to lower pressures suggesting a stable quantum critical phase \cite{Paschen_2021}. 
Such an NFL phase is also argued in a heavy electron system presumably hosting a QSL \cite{Zhao_2019}. 
Charge carriers moving around in QSL phases may not achieve sufficient coherence, losing Fermi degeneracy and resulting in NFL phases. 
Theoretically, the temperature-linear resistivity observed in $\kappa$-HgBr at low pressures is suggested for bosonic charge excitations in a spin-charge separated doped spin liquid \cite{Lee_2006}.

\subsection{Upper critical field and superconducting coherence length : BEC-BCS crossover}
\begin{figure}
	\includegraphics[width=17cm]{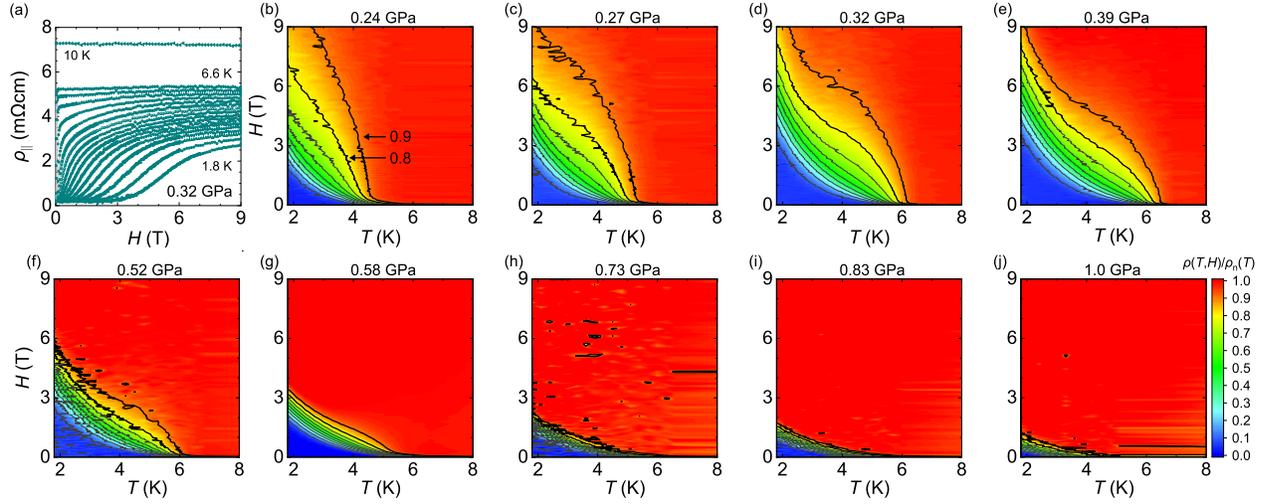}
	\caption{\label{Fig3} (Color online) Resistivity under magnetic field. 
(a) In-plane resistivity as a function of magnetic field, $H$, applied perpendicular to the layers under the pressure of 0.32 GPa. 
The temperature was set to the temperatures between 1.8 and 6.6 K at 0.3 K intervals and to 10.0 K.  
(b)-(j) Contour plots of the normalized in-plane resistivity, $\rho$($T, H$)/$\rho_{\textrm{n}}$($T$), in the temperature ($T$)-magnetic field ($H$) plane for the pressure of 0.24, 0.27, 0.32, 0.39, 0.52, 0.58, 0.73, 0.83 and 1.0 GPa, where $\rho$($T, H$) is the in-plane resistivity at $T$ and $H$ and $\rho_{\textrm{n}}$($T$) is the normal resistivity with the form of ($\rho_{0}$ + $AT^{\alpha}$) fitting the high-field data (see text). 
The solid lines indicate the contours of $\rho$($T, H$)/$\rho_{\textrm{n}}$($T$) = 0.1, 0.2, 0.3, 0.4, 0.5, 0.6, 0.7, 0.8 and 0.9.
}
\end{figure}

Next, we explore how the nature of the superconducting condensate changes upon crossover from the anomalous metal to the ordinary metal by pressure. 
To assess the BEC or BCS nature, we evaluate the in-plane Ginzburg-Landau coherence length $\xi_{\parallel}$, which measures the Cooper-pair size, albeit not strictly in the BEC regime \cite{Engelbrecht_1997}, from the perpendicular upper critical field, $H_{\textrm
{c2}\perp}$. 
To determine $H_{\textrm{c2}\perp}$, we traced the in-plane resistivity upon sweeping the magnetic field normal to the conducting layers shown at fixed temperatures \cite{SM}. 
Figure \ref{Fig3}(a) shows the resistivity evolution upon the destruction of superconductivity by magnetic field at 0.32 GPa as an example (see \cite{SM} for the resistivity data at other pressures, 0.24, 0.27, 0.32, 0.39, 0.52, 0.58, 0.73, 0.83 and 1.0 GPa). 
As the normal state shows no appreciable magnetoresistance as seen in the figure, the field dependence of the resistivity stems from the mobile vortices and/or superconducting fluctuations. 
The resistivity behavior in the magnetic field-temperature ($H$-$T$) plane at each pressure is shown in the contour plot of the reduced resistivity $\rho$($T$, $H$)/$\rho_{\textrm{n}}$($T$), where $\rho_{\textrm{n}}$($T$) is the normal-state resistivity determined above, in Figs. \ref{Fig3}(b)-\ref{Fig3}(j). 
In the blue area, the resistivity vanishes or is small. 
As the magnetic field increases, the system enters the mixed state with mobile vortices, $\rho$($T$, $H$)/$\rho_{\textrm{n}}$($T$) takes finite values (light blue-yellow area), and eventually, the normal-state resistivity is restored, i.e., $\rho$($T$, $H$)/$\rho_{\textrm{n}}$($T$) $\approx$ 1 (red area). 
In the low-pressure regime, there is a wide region with finite values in $\rho$($T$, $H$)/$\rho_{\textrm{n}}$($T$), indicating that mobile vortices or superconducting fluctuations are developed even at high fields. 
In contrast, at high pressures, the superconductivity is fully suppressed by low magnetic fields, and the transition to the normal state is sharp, as in conventional superconductors. 
Overall, superconductivity is robust to magnetic fields at low pressures.

\begin{figure}
	\includegraphics[width=8.6cm]{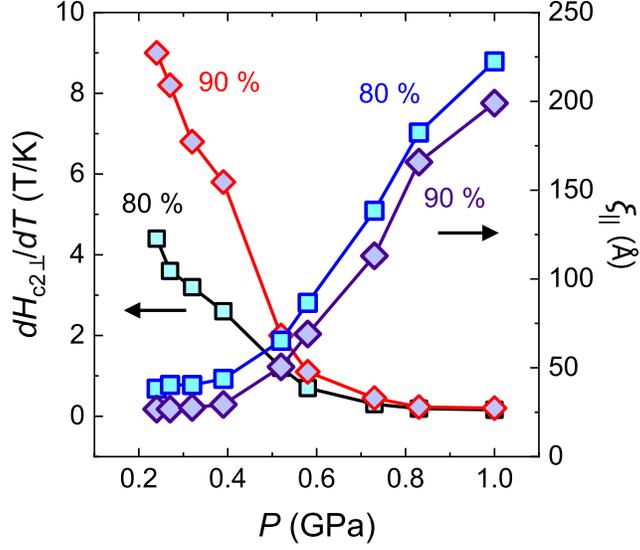}
	\caption{\label{Fig4} (Color online) Upper critical field and Superconducting coherence length. 
The in-plane Ginzburg-Landau coherence length $\xi_{\parallel}$ calculated from the slope of the upper critical field, $dH_{\textrm{c2}}$$_{\perp}$/$dT$ near $T_{\textrm{c}}$, where the $H_{\textrm{c2}}$$_{\perp}$ is determined from the line at $\rho$($T, H$)/$\rho_{\textrm{n}}$($T$) = 0.8, 0.9 (80$\%$ and 90$\%$ of the normal-state resistivity) in Fig. \ref{Fig3}. 
The squares and diamonds mean the values determined for $\rho$($T, H$)/$\rho_{\textrm{n}}$($T$) = 0.8 and 0.9, respectively
}
\end{figure}

\begin{figure}
	\includegraphics[width=8.6cm]{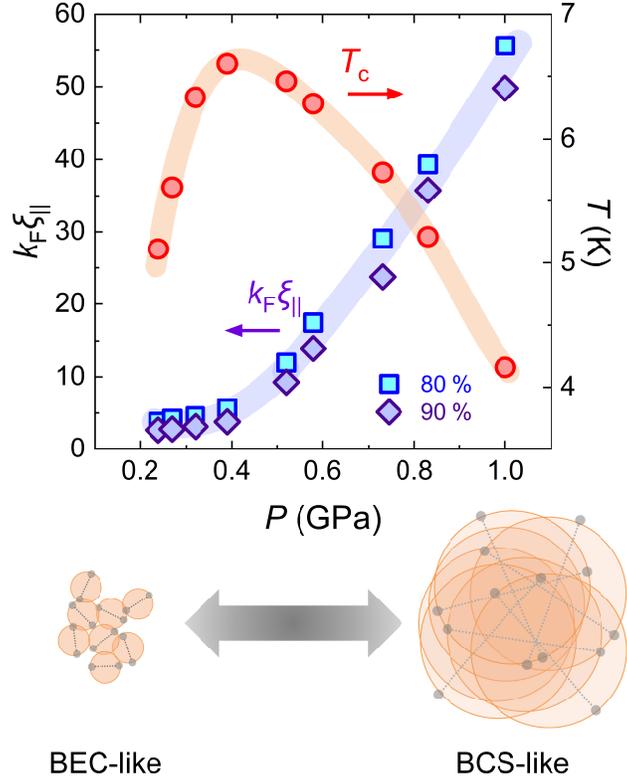}
	\caption{\label{Fig5} (Color online) Pressure dependence of $k_{\textrm{F}}\xi_{\parallel}$ and BEC-BCS crossover. 
Superconducting transition temperature, $T_{\textrm{c}}$, and $k_{\textrm{F}}\xi_{\parallel}$ as a function of temperature, where $k_{\textrm{F}}$ is the Fermi wavenumber determined by the Hall coefficient \cite{Oike_2015}. 
The value of $k_{\textrm{F}}\xi_{\parallel}$ is $\sim$3 at low pressures and increases up to $\sim$50 at 1 GPa with pressure, pointing to a BEC-BCS crossover. 
Schematics drawn underneath the panel depict the BEC- and BCS-like pairs on the low- and high-pressure sides.
 }
\end{figure}

The coherence length $\xi_{\parallel}$ is evaluated from the temperature derivative of $H_{\textrm{c2}\perp}$ near $T_{\textrm{c}}$ through $T_{\textrm{c}}$[$dH_{\textrm{c2}\perp}$($T$)/$dT$]$_{T = T_{\textrm{c}}}$ = $\phi_{0}$/2$\pi$$\xi_{\parallel}^{2}$, where $\phi_{0}$ is the flux quantum. 
In highly two-dimensional layered superconductors, finite resistivity is caused by vortex flow under perpendicular magnetic fields, as observed in this system \cite{Tinkham_2004}; therefore, the definition of $H_{\textrm{c2}\perp}$ is not straightforward. 
In such a case, $H_{\textrm{c2}\perp}$ is better characterized by an initial drop in resistivity. 
Here, we adopt as $H_{\textrm{c2}\perp}$ the 80$\%$ and 90$\%$ transition points ($\rho$($T$, $H$)/$\rho_{\textrm{n}}$($T$) = 0.8 and 0.9), which are indicated by bold contours in Fig. \ref{Fig3}(b)-\ref{Fig3}(j). 
The values of [$dH_{\textrm{c2}\perp}$($T$)/$dT$]$_{T = T_{\textrm{c}}}$ for the the 80$\%$ and 90$\%$ transition lines are shown in Fig. \ref{Fig4}. 
Using these values, we determined the $\xi_{\parallel}$ values, whose pressure dependence is also shown in Fig. \ref{Fig4}. 
At pressures below 0.4 GPa, $\xi_{\parallel}$ $\sim$ 3 nm, whereas, it increases steeply with pressure and reaches $\sim$ 20 nm at 1 GPa. 
Assuming a cylindrical Fermi surface, we evaluated the Fermi wavenumber $k_{\textrm{F}}$ from the Hall coefficient under pressure \cite{SM} and obtained the $k_{\textrm{F}}$$\xi_{\parallel}$ value, an index of the BEC/BCS nature or, roughly speaking, the degree of overlap of Cooper pairs. 
The pressure dependence of $k_{\textrm{F}}$$\xi_{\parallel}$ is shown along with $T_{\textrm{c}}$ in Fig. \ref{Fig5}. 
At high pressures, e.g., 1 GPa, $k_{\textrm{F}}$$\xi_{\parallel}$ reaches $\sim$ 50, pointing to highly overlapping pairs as in the conventional BCS regime, whereas at low pressures, $k_{\textrm{F}}$$\xi_{\parallel}$ is as small as $\sim$ 3; the size of the Cooper pair may become even smaller than the evaluated $\xi_{\parallel}$ value when approaching the BEC regime according to a theoretically suggestion \cite{Engelbrecht_1997}. 
The $k_{\textrm{F}}$$\xi_{\parallel}$ value of the order of unity signifies that the pairing is BEC-like, while the value of several tens points to BCS-like pairing, thus demonstrating that pressure induces a BEC-to-BCS crossover. 
The field-robust vortex liquid state and fluctuations at low pressures are consistent with the short coherence length and small density of the bosonic pairs inherent in BEC-like pairing, as shown theoretically \cite{Adachi_2019}.

\subsection{Nernst effect and superconducting fluctuations}
\begin{figure}
	\includegraphics[width=8.6cm]{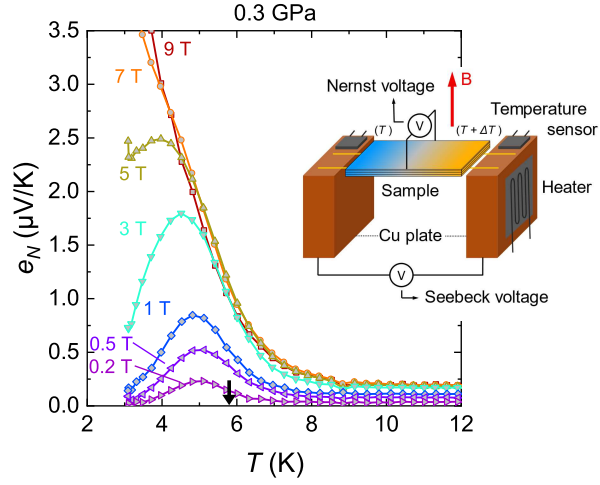}
	\caption{\label{Fig6} (Color online) Temperature dependence of the Nernst signal $e_{N}$ under a pressure of 0.30 GPa. 
The measurements were made at the magnetic fields of 0.2, 0.5, 1, 3, 5, 7 and 9 T applied perpendicular to the conducting layers. 
The arrow indicates the zero-field $T_{\textrm{c}}$ determined by the simultaneously measured thermopower. 
The geometry of the measurement setup is depicted in the inset.
 }
\end{figure}

\begin{figure}
	\includegraphics[width=17cm]{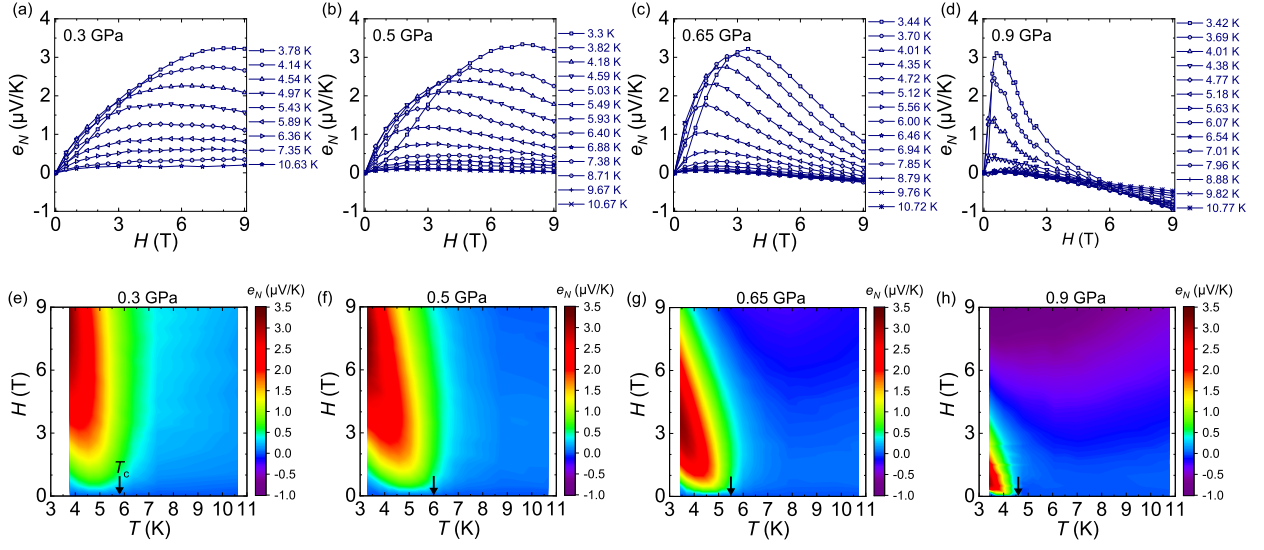}
	\caption{\label{Fig7} (Color online) Magnetic field-temperature profile of the Nernst signal $e_{N}$ under different pressures. 
(a)-(d) Nernst signal $e_{N}$ traced upon sweeping the perpendicular magnetic field at several fixed temperatures under the pressures studied, 0.3, 0.5 0.65 and 0.9 GPa. 
(e)-(h) Contour plots of the Nernst signal eN in the temperature-magnetic field plane at the pressures of 0.3, 0.5 0.65 and 0.9 GPa. 
The red areas indicate hot Nernst regions. The arrows indicate the zero-field $T_{\textrm{c}}$.
 }
\end{figure}

\begin{figure}
	\includegraphics[width=8.6cm]{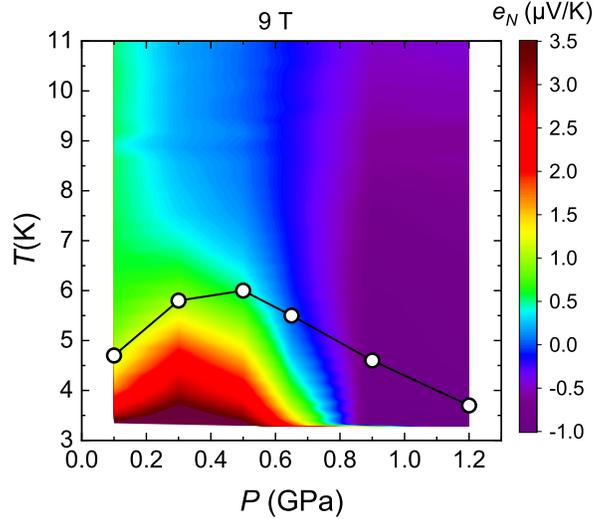}
	\caption{\label{Fig8} (Color online) Temperature-pressure profile of the Nernst signal $e_{N}$. 
The contour plot of the Nernst signal $e_{N}$ at a field of 9T is shown in the temperature-pressure plane. 
The white circles indicate the zero-field $T_{\textrm{c}}$.
 }
\end{figure}

We further verified the preformed nature and magnetic field robustness of the pairing by examining the Nernst effect, in which a thermal flow generated by a temperature gradient $\nabla_{x}T$ under a perpendicular magnetic field induces a transverse electric field $E_{y}$ (see the inset of Fig.\ref{Fig6}). 
This effect, characterized by $e_{N}$ $\equiv$ $\pm$$E_{y}$/$\nabla_{x}T$ (with the sign depending on the field direction), is known to be enhanced by phase fluctuations arising from vortex liquid or preformed pairs in the superconducting state under magnetic fields, as observed in real materials such as cuprate and organic superconductors \cite{Chen_2005, Wang_2006, Kang_2020, Nam_2007, Nam_2013, Behnia_2016}. 
Figure \ref{Fig6} shows the temperature variation of $e_{N}$ at several fixed magnetic fields under a pressure of 0.3 GPa. 
The $e_{N}$ value, which is small in the normal state, starts to increase at temperatures well above $T_{\textrm{c}}$ upon cooling, signaling superconducting preformation; it is remarkable that $e_{N}$ shows a field-insensitive universal temperature variation at high magnetic fields. 
Figures \ref{Fig7}(a)-\ref{Fig7}(d) display the magnetic-field variation of $e_{N}$ at several fixed temperatures under the pressures of 0.3, 0.5, 0.65 and 0.9 GPa. 
It is evident that the magnetic field profile of $e_{N}$ is strongly pressure-dependent; the $e_{N}$ value keeps large even at the maximum field, 9T, under 0.3 GPa whereas under 0.9 GPa it rapidly drops above 1 T. 
To visualize the pressure evolution of the Nernst effect, we displays the contour plots of $e_{N}$ in the temperature-magnetic field plane for each pressure in Figs. \ref{Fig7}(e)-\ref{Fig7}(h). 
At 0.3 GPa, the ridge fields at which $e_{N}$ takes large values are extended vertically and even tend to go beyond the experimental maximum field, 9 T. 
In general, quasi-two-dimensional superconductivity is easily suppressed by the perpendicular magnetic field far less than the Pauli limiting field, which is $\sim$ 10 T at 0.3 GPa, due to the orbital depairing. 
No such symptom up to 9 T and higher evidences unusual robustness of the pairing to the magnetic field, in line with the view of the BEC condensate. 
At higher pressures, the ridge fields become suppressed along with $T_{\textrm{c}}$ and inclined, as in conventional superconductors. 
To see the pressure-temperature profile of $e_{N}$, we make a contour plot of the $T_{\textrm{c}}$ values at the maximum experimental field, 9 T (Fig. \ref{Fig8}), which illustrates that the field-robust Nernst signal and its persistence at high temperatures far beyond $T_{\textrm{c}}$ is specific to the low-pressure phase. 
This is the Nernst behavior typical of BEC-like condensates; the effect of preformed pairs is expected to set in at higher temperatures in the Nernst signal than in the resistivity \cite{Chen_2005}, consistent with the present observations. 
The systematic change of the Nernst profile with pressure lends support to the BEC-BCS crossover of the superconducting condensate.

It is reported that the $^{13}$C NMR relaxation rate divided by temperature, 1/$T_{1}T$, shows a decrease below 7-9 K \cite{Eto_2010} despite of $T_{\textrm{c}}$ $\sim$ 4.2 K
at ambient pressure and the decrease becomes less prominent at higher pressures. This behavior is a likely manifestation of the preformed pairs in the BEC-like regime at ambient pressure and its crossover to the BCS regime at high pressures. We also note that the recent torque measurements suggest the fluctuating superconductivity that sets in at around 7 K at ambient pressure, implying the preformation of the Cooper pairs \cite{Imajo_2021}.

\section{CONCLUDING REMARKS}
The electron transport and Nernst effect in the doped triangular system, $\kappa$-HgBr, revealed that the nature of the superconductivity crosses over from BEC-like to BCS pairing with increasing pressure.
The present BEC-BCS crossover is associated with an NFL-FL crossover, at which the Coulombic repulsive energy relative to the band width exceeds a critical value for a Mott transition that would occur unless doped. 
To explore the nature of fermionic pairing that encompasses the BEC and BCS regimes, it is crucial to vary the control parameter linked to the interparticle interaction across some critical value at which the nature of the interaction dramatically changes.
In cold fermionic atoms, the controlling parameter is the magnetic field, which varies interatomic interactions across a critical value through the Feshbach resonance. 
In the present case of interacting electrons, pressure is a parameter that varies the Coulombic interactions across the Mott critical value at which the double occupancies of carriers are critically allowed or forbidden \cite{Hebert_2015}. 
This Mott-driven BEC-BCS crossover is a novel addition to the physics of the pairing instability of interacting fermions.

\begin{acknowledgements}
	We thank M. Ogata for his fruitful discussions. 
	This work was supported by Japan Society for the Promotion of Science (grant Nos. 18H05225, 19H01846, 20K20894 and 20KK0060). 
	Several experiments were performed using facilities of the Cryogenic Research Center, the University of Tokyo.
\end{acknowledgements}

\begin{appendix}

\appendix*
\section{METHODS}
\subsection{Measurements of resistivity}
The in-plane resistivity was measured by the conventional four-probe method; four gold wires were attached on the planar surface of a crystal with carbon paste. 
The Quantum Design Physical Property Measurement System (PPMS) was used for the resistivity measurements. 
To investigate the perpendicular upper critical field, $H_{c2}$$_{\perp}$, we traced the resistivity with the magnetic field applied perpendicular to the conducting layers up to 9 T. 
In organic conductors, the rapid cooling often causes unwanted crystal cracking and/or conformational disorder of terminal ethylene groups in BEDT-TTF. 
To avoid suffering from these, the cooling rate was kept below 0.5 K/min.

\subsection{Measurements of the Nernst effect}
The Nernst voltage, $e_{N}$, was generated in a direction perpendicular to both the directions of an externally created temperature gradient and an applied magnetic field (see the inset of Fig. \ref{Fig6}). 
The magnetic field was applied perpendicular to the conducting layers. The temperature difference across the crystal, $\Delta$$T$, along the two-dimensional plane was kept less than 0.5 K in the field-sweep measurements and less than $T$/10 in the temperature-sweep measurements. 
The temperatures were measured with the Cernox sensors (Lake Shore Cryotronics, Inc.) attached on the cold and hot side of the crystal. 
The Nernst-effect measurements were performed under several pressures and the Cernox sensors were calibrated every time pressure is changed, using a reference thermometer. 
To determine the Nernst voltage, $e_{N}$, the transverse voltage $V$ was measured under both positive and negative fields (+$H$ and -$H$), and then the $e_{N}$ was obtained by antisymmetrizing the two voltages by $e_{N}$($H$) = ($V$(+$H$)-$V$(-$H)$)/2. 
Through this measurement, PPMS was used as a controller of magnetic field and temperature; the temperature gradient in the sample was generated by a heater attached on one side of the crystal and the sample temperature was defined by the mean value of the temperatures of the hot and cold sides. 
As in the resistivity measurements, the cooling rate was kept below 0.5 K/min. 
The thermopower was also measured along with the Nernst voltage for the identical sample in order to know the $T_{\textrm{c}}$ of the sample used.

\subsection{Pressurization}
To apply hydrostatic pressures to the sample, a dual structured clamp-type pressure cell made from BeCu and NiCrAl cylinders was used with the Daphne 7373 oil as a pressure-transmitting medium. 
The pressure values quoted in this article are the internal pressures at low temperatures determined with the manganin pressure gauge or the tin manometer.
According to Ref. \cite{Murata_2007}, the clumped pressure gradually decreases by 1.5-2.0 kbar on cooling from 300 K to 50 K and is nearly constant below that.

\end{appendix}

$\\$

$^{\ast}$Corresponding author, e-mail: kanoda@ap.t.u-tokyo.ac.jp

$^{\dagger}$These authors contributed equally to this work

\vspace{2ex}\vspace{2ex}

\centerline{\textbf{\large{Supplemental Material for the Papaer}}}

\centerline{\textbf{\large{"Mott-driven BEC-BCS crossover in a doped spin liquid}}}

\centerline{\textbf{\large{candidate, $\kappa$-(BEDT-TTF)$_{4}$Hg$_{2.89}$Br$_{8}$"}}}

\vspace{2ex}

\textbf{Crystal structure and band filling in $\kappa$-(BEDT-TTF)$_{4}$Hg$_{2.89}$Br$_{8}$}

The family of organic charge-transfer salts, $\kappa$-(BEDT-TTF)$_{2}X$, where BEDT-TTF and $X$ stand for bis(ethylenedithio)tetrathiafulvalene and anions, respectively, have layered structures composed of BEDT-TTF conducting layers and insulating anion layers, both of which have monoclinic sublattices with the space group $C2$/$c$ and $C2$/$m$, respectively \cite{Li_1998}. 
In the conducting layers, the BEDT-TTF molecules form dimers, which constitute a quasi-triangular lattice characterized by two kinds of transfer integrals, $t$ and $t^{\prime}$, between the adjacent dimer molecular orbitals (Fig. 1(b)). 
According to the molecular orbital calculations, the ratio, $t^{\prime}$/$t$, in $\kappa$-(BEDT-TTF)$_{4}$Hg$_{2.89}$Br$_{8}$ is 1.02 \cite{Shimizu_2011}, which is compared to the value in the Mott-insulating spin liquid candidate, $\kappa$-(BEDT-TTF)$_{2}$Cu$_{2}$(CN)$_{3}$, 1.06 \cite{Komatsu_1996} (0.83 \cite{Kandpal_2009} and 0.83-0.99 \cite{Koretsune_2014} according to the first-principles calculations).

Most of $\kappa$-(BEDT-TTF)$_{2}X$ compounds have half-filled bands, or equivalently, one hole carrier per one dimer because the valence of the anion $X$ is -1.
 However, the band filling in $\kappa$-(BEDT-TTF)$_{4}$Hg$_{2.89}$Br$_{8}$ deviates from a half by 0.055 assuming Hg$^{+2}$ and Br$^{-1}$, because of the nonstoichiometry of Hg, whose sublattice is incommensurate along the c-axis against those of BEDT-TTF and Br; if the composition of Hg were 3.0, the band filling would be a half. 
The compositional ratio of 4.0 : 2.89 : 8.0 cannot be varied because the incommensurability precisely determined by x-ray diffraction \cite{Li_1998, Lyubovskaya_1987} results from chemistry between BEDT-TTF, Hg and Br. 
Such non-stoichiometry occurs in other Hg-containing compounds as well \cite{Gillespie_1985}.
The band filling slightly deviated from a half is evidenced by the valence of BEDT-TTF determined by the Raman spectroscopy \cite{Yamamoto_2005}. 
This band-filling deviation from a half corresponds to a 11$\%$ hole doping to the half-filled band.

\vspace{2ex}

\textbf{Determination of the upper critical field} 

\begin{figure}
	\includegraphics[width=15cm]{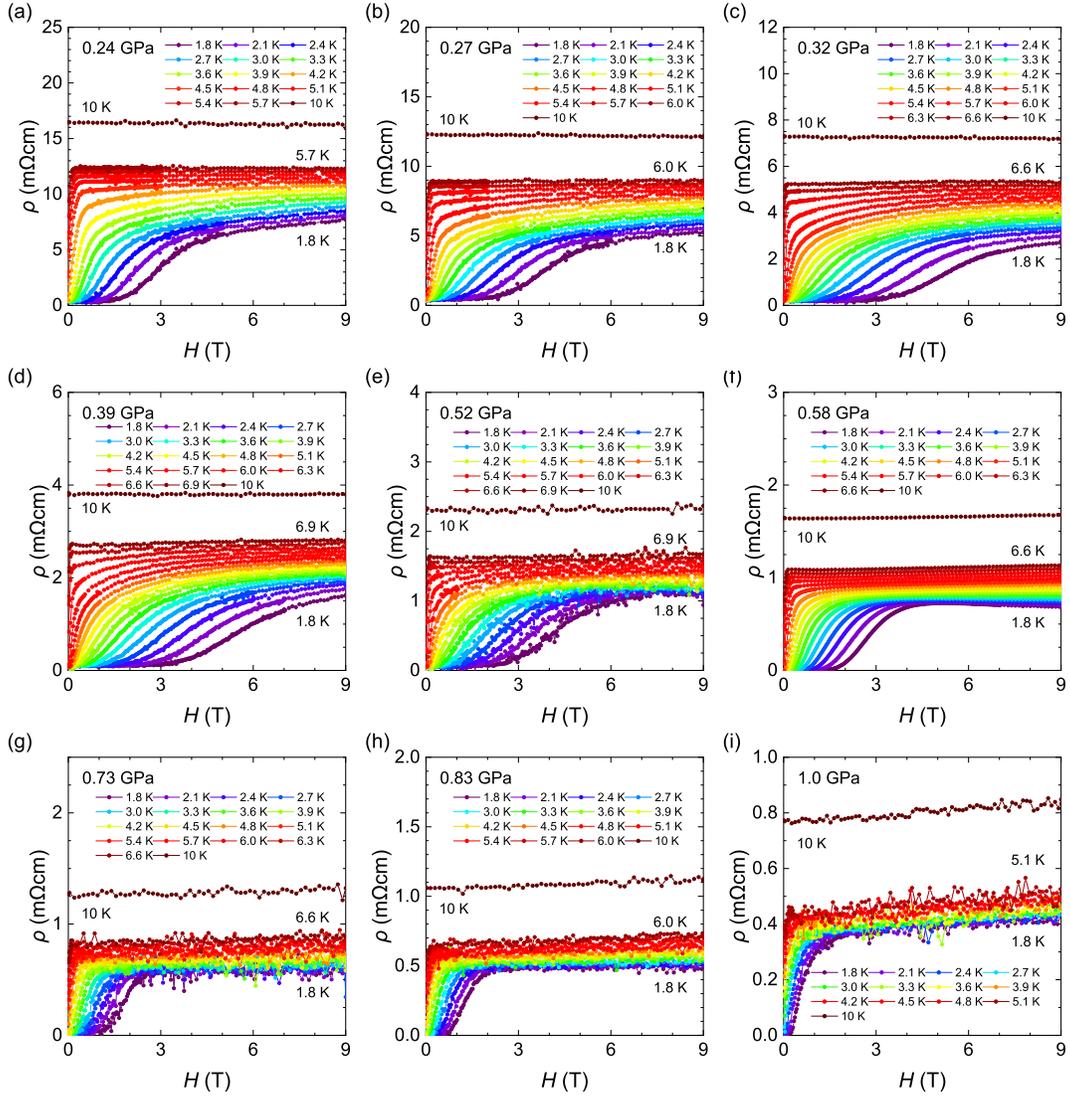}
	\caption{\label{FigS1}(Color online) 
Magnetic field dependence of the in-plane resistivities under pressure.
The magnetic field was applied perpendicular to the conducting layer. 
Temperature was fixed at intervals of 0.3 K below $T_{\textrm{c}}$. 
For reference data at a high temperature well above $T_{\textrm{c}}$, the resistivity curves at 10 K were measured.
}
\end{figure}

The upper critical field is usually defined as the magnetic field at which the vanishingly small resistivity transitions to the normal-state one. 
In highly two-dimensional layered superconductors like the present material, however, resistivity is induced by mobile vortices or, broadly speaking, superconducting phase fluctuations under perpendicular field even when the amplitude of the superconducting order parameter is developed. Figure \ref{FigS1} shows the magnetic field dependence of the in-plane resistivities at each pressure, where the magnetic field was applied normal to the conducting plane. 
As expected, the resistive transition against field variation is gradual, in particular, for low pressures. 
Therefore, we defined the perpendicular upper critical filed $H_{\textrm{c2}}$$_{\perp}$ by the field at which the resistivity drops by 10$\%$ and 20$\%$ from the normal-state value, $\rho_{\textrm{n}}(T)$, of the form, $\rho_{\textrm{n}}(T)$ = $\rho_{0}$ + $AT^{\alpha}$, fitting the normal state resistivity; namely, $H_{\textrm{c2}}$$_{\perp}$ is defined in two ways as the fields of $\rho(T, H)$/$\rho_{\textrm{n}}(T)$ = 0.8 and 0.9, where $\rho(T, H)$ is the resistivity at temperature $T$ and magnetic field $H$. 
The contour curves of $\rho(T, H)$/$\rho_{\textrm{n}}(T)$ = 0.8 and 0.9 are shown by the bold lines in Fig. 3(b)-(j).

\vspace{2ex}

\textbf{Evaluation of $k_{\textrm{F}}\xi_{\parallel}$}

\begin{figure}
	\includegraphics[width=15cm]{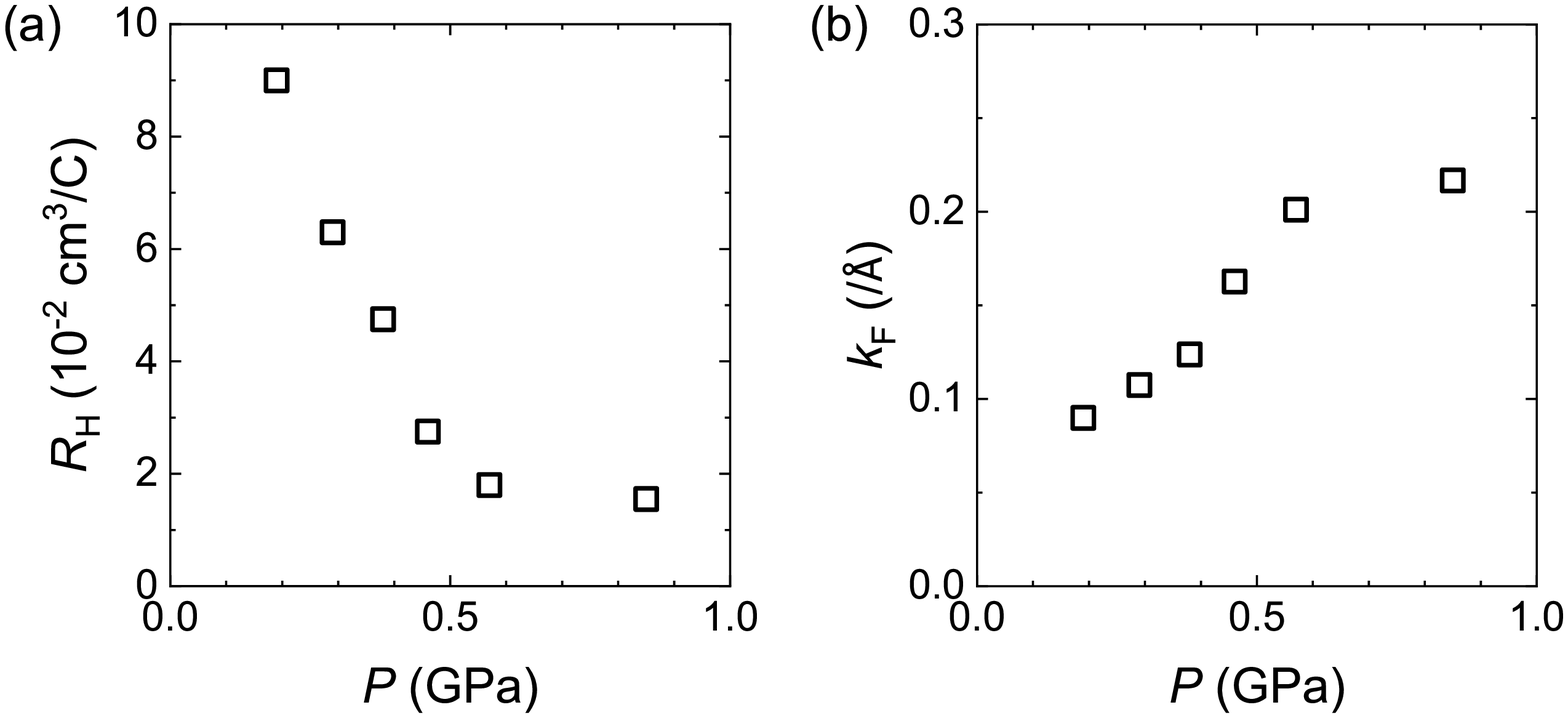}
	\caption{\label{FigS2} (Color online) 
Hall coefficient of $\kappa$-HgBr at 10 K \cite{Oike_2015} and the estimate of $k_{\textrm{F}}$. 
(a) Pressure dependence of the Hall coefficient at 10 K. The values are taken from Ref. \cite{Oike_2015}. (b) $k_{\textrm{F}}$ values deduced from the Hall coefficient are shown in Fig. \ref{FigS2}(a) on the assumption of the relation, $R_{\textrm{H}}$ = 1/$ne$, and the two-dimensional cylindrical Fermi surface.
}
\end{figure}

The in-plane coherence length $\xi_{\parallel}$ was calculated from the slope of $H_{\textrm{c2}}$$_{\perp}$ near $T_{\textrm{c}}$, $dH_{\textrm{c2}}$$_{\perp}$ /$dT$, through the formula, $\xi_{\parallel}$ = $\sqrt{\frac{\phi_{0}}{2\pi{T}_{\textrm{c}}|dH_{\textrm{c2}\perp}/dT|}}$
The experimental $dH_{\textrm{c2}}$$_{\perp}$ /$dT$ and $T_{\textrm{c}}$ values are shown in Fig. 4 and Fig. 5 in main text. The Fermi wavenumber, $k_{\textrm{F}}$, was estimated from the Hall coefficient data \cite{Oike_2015} using the relation $R_{\textrm{H}}$ = 1/$ne$ and assuming the two-dimensional cylindrical Fermi surface, where $n$ is the carrier density and $e$ is the elementary charge. 
The pressure dependences of the $R_{\textrm{H}}$ value at 10 K \cite{Oike_2015} and the calculated $k_{\textrm{F}}$ values are shown in Fig. \ref{FigS2}: the $k_{\textrm{F}}$ values at pressures studied here were obtained by interpolating the points in Fig. \ref{FigS2}(b).

\newpage

\newpage

\newpage

\newpage

\newpage

\newpage

\newpage

\newpage

\newpage

\newpage

\newpage

\end{document}